\documentclass[showpacs, twocolumn]{revtex4}
\usepackage{graphicx}

\newcommand{\x}{{\bf r}}
\newcommand{\K}{{\bf k}}

\begin{document}

\title{Pair correlations of scattered atoms from two colliding
Bose-Einstein Condensates: Perturbative Approach.}

\author{J. Chwede\'nczuk$^{1}$, P. Zi\'n$^{2}$, M. Trippenbach$^{1,2}$, A. Perrin$^{3,4}$, V. Leung$^{3}$, D. Boiron$^{3}$
and C. I. Westbrook$^{3}$}

\affiliation{$^1$Institute of Theoretical Physics, Physics Department,
Warsaw University, Ho\.{z}a 69, PL-00-681 Warsaw,
Poland\\$^2$The Andrzej So{\l}tan Institute for Nuclear
Studies, Warsaw University, Ho\.{z}a 69, PL-00-681 Warsaw,
Poland\\$^3$Laboratoire Charles Fabry de l`Institut d`Optique,
CNRS, Univ Paris-Sud, Campus Polytechnique, RD128, 91127 Palaiseau
cedex, France\\$^4$Atominstitut der \"{O}sterreichischen
Universit\"{a}ten, TU-Wien, Stadionallee 2, A-1020 Vienna,
Austria}

\begin{abstract}
We apply an analytical model for anisotropic, colliding
Bose-Einstein condensates in a spontaneous four wave mixing geometry to
evaluate the second order correlation function of the field of scattered
atoms. Our approach uses quantized scattering modes and the equivalent of a
classical, undepleted pump approximation. Results to lowest order in perturbation
theory are compared with a recent experiment and with other theoretical
approaches. 
\end{abstract}

 \maketitle

\section{Introduction}

The analog of correlated photon pair production \cite{burnham:70} has recently
been demonstrated using atoms. Both molecular dissociation \cite{greiner:05} and
four wave mixing of deBroglie waves \cite{perrin:07} have shown correlation
peaks. As in quantum optics, such atom pairs lend themselves to
investigations into non-classical correlation phenomena such as entanglement
of massive particles \cite{duan:00,pu:00,opatrny:01,kheruntsyan:05} and spontaneous directionality or superradiant
effects \cite{pu:00,vardi:02}.  From the point of view of the outgoing atoms, the underlying
physics is very similar and thus theoretical descriptions should be
applicable to both processes. The experiment using four wave mixing of
metastable helium atoms in particular has yielded detailed information about
the atomic pair correlations. Efforts to treat the experimental situations
are therefore highly desirable.

Theoretically, the description of condensate collisions in the
spontaneous scattering regime requires a formulation that extends
beyond the mean-field model \cite{bach:02,yurovsky:02}. 
In previous work on spherical
Gaussian wave packets, within perturbative approach, we have given
analytical formulas for the correlation functions \cite{zin:05, zin:06}. 

In this paper we extend  our method to anisotropic condensates to give an analytic description 
of  the correlation properties of spontaneously emitted atom pairs in a geometry much closer to and in
good agreement with the experiment \cite{perrin:07}.
Numerical approaches using truncated Wigner method \cite{norrie:05,norrie:06} 
and positive-P method \cite{savage:06,deuar:07,perrin:08} 
have also been used, in particular to give insight into the stimulation 
regime where bosonic enhancement comes to play.

Here, we use the model of colliding condensates to examine two types of correlations. First we shall focus
on atom pairs originating from the same two body scattering event. These consequently have nearly opposite momenta.
Thus we analyze the opposite-momenta correlations of atom pairs. Second, we examine two body correlations between atoms 
scattered with nearly collinear momenta, a manifestation of the Hanbury
Brown-Twiss (HBT) effect \cite{zin:05, savage:06, deuar:07,
molmer:07}. 
In both cases, the demonstration of a two particle correlation requires
a measurement of the
conditional probability of detecting a particle at position $\x_1$
given that a particle was detected at $\x_2$. This
probability is proportional to the second order correlation
function $G^{(2)}(\x_1,\x_2)$ of the field $\hat\delta$ of atoms,
i.e.
\begin{eqnarray*}
  G^{(2)}(\x_1,\x_2)={\langle\hat\delta^\dagger(\x_1)\hat\delta^\dagger(\x_2)\hat\delta(\x_2)\hat\delta(\x_1)\rangle}.
\end{eqnarray*}
We shall pay particular attention to
correlations in momentum space and compare these results with experimental data of \cite{perrin:07}.
A careful comparison of a numerical treatment based on the positive-P method \cite{perrin:08}
with the experiment \cite{perrin:07}  indicated reasonable agreement, but one of the limitations of 
the method, the short collision duration which could be simulated, left
some unresolved questions.
In particular, energy conservation is a less stringent constraint for
short collision times, and thus one can wonder about the 
role this constraint plays in the experiment.
The treatment given here is not subject to this limitation
and also agrees fairly well with the experiment for most of the 
experimentally accessible observables. 
One observable quantity however,  the averaged width of the collinear correlation function in
a direction orthogonal to the symmetry axis, disagrees with the 
experiment and with Ref. \cite{perrin:08}. 
In our treatment, it is precisely the requirement of energy conservation
that is at the origin of the difference. 
At the end of the paper we shall discuss possible explanations of this discrepancy.

\begin{figure}[htb]
  \centering
  \includegraphics[scale=0.4, angle=0]{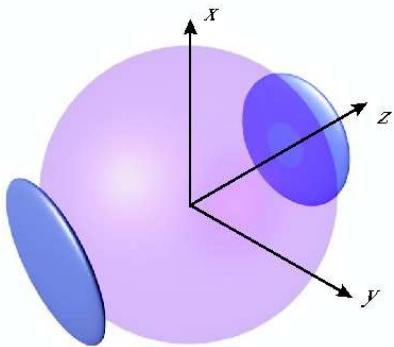}
  \caption{Velocity space representation of the pair production
    experiment.  Raman pulses generate counter-propagating condensates
    which collide and expand into disk-shaped clouds along the
    $z$-axis. Atoms scattered during the collision expand to form a
    spherical shell of correlated pairs.  Note that the orientation of
    the axes in this article differs from Ref. \cite{perrin:07} }
  \label{mcp}
\end{figure}

Let us first describe the
experiment in which a collision of two Bose-Einstein
condensates of metastable helium produces
a cloud of scattered atoms. A condensate of approximately 10$^5$
He$^*$ atoms is created in a cigar-shaped magnetic trap with axial and radial trapping frequencies
of $\omega_z= 2 \pi \times$
47~Hz and $\omega_r= 2 \pi \times$ 1150~Hz respectively. Three
laser beams are used to transfer the atoms into two
counter-propagating wave-packets by a Raman process, with a
transfer efficiency of about 60\%. As the wave-packets
counter-propagate with a relative velocity of
$2v_{rec}=18.4$~cm/s, atoms from the two clouds collide via
$s$-wave scattering, populating a 
spherical shell in momentum space often referred to as the ``halo"
~\cite{chikkatur:00, gibble:95, katz:05}. 
In the experiment, about 5\% of the atoms are scattered.
In addition to
splitting the condensate, the Raman transition transfers the
atoms into an untrapped magnetic sub-state.  The transferred atoms
thus expand freely, falling onto a micro-channel plate (MCP)
detector that allows the three-dimensional reconstruction of the
position of single atoms with an estimated efficiency of
10\%~\cite{schellekens:05,jeltes:07}. Knowing the positions of
individual atoms, the initial momenta, and the second order momentum 
correlation function of the
cloud of scattered particles can be computed. 
The precision of the
measurement is limited by the finite  resolution of
the MCP. This factor will be taken into account in our
comparison between the theoretical estimates and the experimental
results.
\section{Model for scattering}
To make the comparison, we introduce a simplified model for
atom scattering during a collision of two Bose-Einstein condensate
wave-packets. In this model we assume that two counter-propagating
wave-packets constitute a classical undepleted source for the
process of scattering. This concept is introduced in analogy to
examples in quantum optics, where a strong coherent laser field is
treated as a classical wave and its depletion is neglected \cite{ScullyQO}. 
We shall simplify the model further on. 
Since we assume that the two
colliding condensates remain undepleted, the
population of the $\hat\delta$ field of scattered atoms should be
small, as compared to the number of atoms in the condensates. In
such a regime, a Bogoliubov approximation is often used \cite{Lif:80, Ohberg:97}, leading
to linearized equations of motion for the quantum fields. In our
case, the $\hat\delta$ field of scattered atoms satisfies the
Heisenberg equation (for details of the derivation, see
\cite{zin:05, zin:06})
\begin{equation}\label{heis}
i\hbar\partial_t\hat\delta(\x, t)= -\frac{\hbar^2\nabla^2}{2m}
\hat\delta(\x, t)+2g\psi_Q(\x, t)\psi_{-Q}(\x,
t)\hat{\delta}^\dagger(\x, t).
\end{equation}
Here $\psi_{\pm Q}(\x,t)$ is the c-number wave-function of the
colliding condensates with mean momentum per atom equal to $
\pm\hbar Q$. Moreover, the coupling constant $g = \frac{4\pi
\hbar^2 a}{m}$ is related to the atomic mass $m$ and
$s$-wave scattering length $a$ of He$^*$.

To permit analytic calculations, we model the
condensate wave functions $\psi_{\pm Q}(x,y,z,t)$  as Gaussians:
\begin{eqnarray}\label{pump}
  &&\psi_{\pm Q}=\sqrt\frac{N}{2\pi^{3/2}\sigma_r^2\sigma_z}\exp\left(\mp iQz-\frac{i\hbar Q^2
    t}{2m}\right)\nonumber\\&&\times\exp\left(-\frac{1}{2\sigma_r^2}
  (x^2+y^2)-\frac{1}{2\sigma_z^2}(z\mp\frac{ \hbar Qt}{m})^2\right),
\end{eqnarray}
where $N$ is the total number of particles in both wave-packets. The radial ($\sigma_r$)
and axial ($\sigma_z$) width of the Gaussians are extracted from the initial condensate 
wave-function $\Psi_0$ which is calculated numerically from the Gross-Pitaevski equation using an
imaginary time method. In practice we fit $\int dv_x\int dv_y|\Psi_0(\bf{v})|^2$ with
a Gaussian function $\propto\exp(-v_z^2/\chi_z^2)$ and then use $\sigma_z=\hbar/(m\; \chi_z)$. 
We define $\sigma_r$ similarly. Here, for simplicity, we neglect the spread of the condensates
during the collision. This assumption seems reasonable because most of the atom collisions take place 
before the two clouds have had time to expand.

It is useful to change variables and rescale the field operator
\begin{eqnarray*}\nonumber
  \frac{\hbar Q}{m \sigma_z}t \rightarrow t  \ \ \ \ \ \ \  \x
  /\sigma_z \rightarrow \x \ \ \ \ \ \ \ \frac{1}{\sigma_z^{3/2}}
  \hat \delta (\x,t) \rightarrow \hat \delta (\x,t)
\end{eqnarray*}
which simplifies the equation of motion (\ref{heis}), i.e.
\begin{equation}\label{main}
i\beta\partial_t\hat \delta(\x,t)=-\frac{1}{2}\nabla^2 \hat
 \delta(\x,t)+ \alpha    e^{-\frac{x^2+y^2}{\gamma^2}-z^2}e^{-i\beta t-t^2}\hat\delta^\dagger(\x,t),
\end{equation}
where $\alpha=\frac{4Na\sigma_z}{\sigma_r^2\sqrt\pi}$,
$\beta=Q\sigma_z$, $\gamma=\frac{\sigma_r}{\sigma_z}$.

The condensate density in momentum space then reads,
\begin{equation}
|\Psi_0({\bf k})|^2=\frac{N\beta^3}{\sqrt{\pi^3}\,\gamma^2}\exp\left[-\beta^2(k_z^2+\gamma^2k_r^2)\right]
\label{fct_BEC}
\end{equation}
The three parameters $\alpha,\beta$ and $\gamma$ fully determine the dynamics of the field
of scattered atoms. For $N=10^5$ and $\hbar Q=m v_{\mathrm{rec}}$ we have
$\alpha=1053$, $\beta = 227$, and $\gamma = 0.05$. 

We also find  $\chi_z=0.004\ v_{\mathrm{rec}}$, $\sigma_z=39\ \mu$m, $\chi_r=0.0870 \ v_{\mathrm{rec}}$ 
and $\sigma_r=2\ \mu$m.
The parameter $\alpha$ is a measure of the strength of the interactions between particles. 
As such, it governs the fraction of atoms scattered into the halo. 
As a consistency check, in Appendix A we give an alternate estimate of $\alpha$ in the experiment 
using the observed fraction of scattered atoms.

In Section \ref{sec2} we derive an analytical expression for the second order
correlation function in the perturbative regime. It is still an
open question whether, for these parameters, the perturbative approach
applies. We tackle this issue after the evaluation of the
$G^{(2)}$ function is Section \ref{sec2_ap}. In Section \ref{sec_exp} we compare the perturbative results with
the experimental data of \cite{perrin:07}.

\section{Derivation of $G^{(2)}$ in perturbative regime}
\label{sec2}
We shall begin the analytical calculations with a definition of
the Fourier transform of the $\hat\delta$ operator
\begin{equation}
  \hat\delta(\x,t)=\left(\frac{\beta}{2\pi}\right)^{3/2} \int
  d\K\ e^{i\beta\K\x-i\beta k^2
    t/2}\hat\delta(\K,t).\label{trans}
\end{equation}
This particular form of Fourier transformation ``incorporates''
the free evolution of the field. Substitution of Eq.(\ref{trans})
into Eq.(\ref{main}) gives
\begin{eqnarray*}
  &&\partial_t\hat\delta(\K,t)=\mathcal{A}e^{-i\beta t}e^{-t^2}\int d\K'
  e^{\frac{i\beta}{2}(k^2+k'^2)t}\\
  &&\exp\left(
  -\frac{\gamma^2\beta^2}{4}(\K_r+\K'_r)^2-\frac{\beta^2}{4}(k_z+k'_z)^2\right)\hat\delta^\dagger(\K',t),
  \nonumber
\end{eqnarray*}
where $\mathcal{A}=-i\frac{\alpha\beta^2\gamma^2}{8\pi^{3/2}}$,
$\K_r=k_x{\bf e}_x+k_y{\bf e}_y$, and ${\bf e}_i$ is a unit vector
in $i$ direction. The above can be integrated formally, giving
\begin{eqnarray*}
  &&\hat\delta(\K,t)=\mathcal{A}\int_0^t \mbox{d}\tau  \, e^{-i\beta
    \tau} e^{-\tau^2}
  \int d\K' \, e^{\frac{i\beta}{2}(k^2+k'^2)\tau}\\
  &&\exp\left(
  -\frac{\gamma^2\beta^2}{4}(\K_r+\K'_r)^2-\frac{\beta^2}{4}(k_z+k'_z)^2\right)\hat\delta^\dagger(\K',\tau).\nonumber
\end{eqnarray*}

Since in the Heisenberg picture the scattered field remains in its initial vacuum state and the evolution
of the $\hat\delta$ field is linear, the second order
correlation function $G^{(2)}(\K_1,\K_2)$ decomposes into
\begin{eqnarray}
  &&G^{(2)}(\K_1,\K_2;t)=\langle \hat \delta^\dagger (\K_1,t) \hat\delta^\dagger (\K_2,t)\hat \delta (\K_2,t)
  \hat \delta (\K_1,t) \rangle\nonumber\\
  &&=G^{(1)}(\K_1,\K_1;t)\cdot G^{(1)}(\K_2,\K_2;t) + \left| G^{(1)}(\K_1,\K_2;t)
  \right|^2\nonumber\\
  &&+ \left|  M(\K_1,\K_2;t) \right|^2\label{korelacja}
\end{eqnarray}
where $M(\K_1,\K_2;t)=\langle \hat \delta (\K_1,t)  \hat\delta
(\K_2,t) \rangle$ is the anomalous density and
$G^{(1)}(\K_1,\K_2;t)=\langle \hat \delta^\dagger (\K_1,t)
\hat\delta (\K_2,t)\rangle$ is the first order correlation
function. Below we calculate these two functions in the lowest order and for a time $t=\infty$ because all the measurements are made long after the collision has finished. We expand
$\hat\delta$ in a series of perturbative
solutions,
\begin{eqnarray*}
  \hat\delta(\K,t=\infty)=\sum_{i=0}^\infty\hat\delta^{(i)}(\K).\nonumber
\end{eqnarray*}
where in the lowest order we get
\begin{eqnarray}\label{pert_delt}
  &&\hat\delta^{(1)}(\K)=\mathcal{A}\int_0^\infty \mbox{d}\tau  \,e^{-i\beta \tau} e^{-\tau^2}
  \int d\K' \, e^{\frac{i\beta}{2}(k^2+k'^2)\tau}\\
  &&\exp\left(
  -\frac{\gamma^2\beta^2}{4}(\K_r+\K'_r)^2-\frac{\beta^2}{4}(k_z+k'_z)^2\right)\hat\delta^{(0)\dagger}(\K').\nonumber
\end{eqnarray}


\subsection{Anomalous density: $\K_1\simeq -\K_2$ correlations}

The anomalous density in the first order is expressed by
\begin{eqnarray*}
  M(\K_1,\K_2)=\langle \hat \delta^{(0)} (\K_1)
  \hat\delta^{(1)} (\K_2) \rangle.\nonumber
\end{eqnarray*}
Using Eq.(\ref{pert_delt}) we get
\begin{eqnarray*}
  &&M(\K_1,\K_2)=\mathcal{A}\exp\left(-\frac{\gamma^2\beta^2}{4}(\K_{1,r}+\K_{2,r})^2\right)\\
  &&\times\exp\left(-\frac{\beta^2}{4}(k_{1,z}+k_{1,z})^2\right)
  \int_0^\infty \mbox{d}\tau \, \exp\left(-i\beta\Delta\tau-\tau^2\right)\nonumber,
\end{eqnarray*}
where $\Delta = \beta \left(1 - \frac{k_1^2+k_2^2}{2} \right) $.
This gives
\begin{eqnarray}\label{m_final}
  &&M(\K_1,\K_2)= -i\frac{\alpha\beta^2\gamma^2}{16\pi} \exp\left(-\frac{\beta^2}{4}(k_{1,z}+k_{2,z})^2\right)\\
  && \exp\left(
  -\frac{\gamma^2\beta^2}{4}(\K_{1,r}+\K_{2,r})^2 - \frac{ \Delta^2}{4}\right)
  \left( 1 - \mbox{erf} \left( \frac{i \Delta}{2} \right)  \right).\nonumber
\end{eqnarray}
This expression shows that the anomalous density describes the
correlations of atoms with opposite momenta. In other words, it is
non-negligible only when $\K_1\simeq -\K_2$. If this condition is
not satisfied, the exponential functions drop quickly. Comparing this expression to Eq.(\ref{fct_BEC}), we find that the widths of the anomalous density have the same anisotropy and are two times larger than the condensate density.
Moreover, this expression shows that this function is also non-negligible only for $\Delta\lesssim 1$. As $\beta$
is large, $\Delta\sim 1$ only when $k_1\simeq1$ and $k_2\simeq1$.
This requirement expresses the conservation of energy in the collision of two atoms.

\subsection{First order correlation function: $\K_1\simeq \K_2$ correlations}

In the lowest order we have
\begin{eqnarray*}
  G^{(1)}(\K_1,\K_2)=\langle\hat\delta^{(1)\dagger}(\K_1)\hat\delta^{(1)}(\K_2)\rangle.\nonumber
\end{eqnarray*}
Using Eq.(\ref{pert_delt}) and
$\langle\hat\delta^{(0)}(\K_1)\hat\delta^{(0)\dagger}(\K_2)\rangle=\delta^{(3)}(\K_1-\K_2)$
we get
\begin{widetext}
\begin{eqnarray*}
  && G^{(1)}(\K_1,\K_2)=|\mathcal{A}|^2\int_0^\infty \mbox{d} \tau\int_0^\infty \mbox{d}\tau' \,
  \exp\left[-\tau^2-\tau'^2+i\beta(\tau-\tau')\right]\int d\K \,
  \exp\left[-\frac{\gamma^2\beta^2}{4}\left((\K_{1,r}+\K_r)^2+(\K_{2,r}+\K_r)^2\right)\right]\nonumber\\
  &&\times\exp\left[-\frac{\beta^2}{4}\left((k_{1,z}+k_z)^2+(k_{2,z}+k_z)^2\right)\right]
  \exp\left[i\frac{\beta}{2}(k^2+k_2^2)\tau'-i\frac{\beta}{2}(k^2+k_1^2)\tau\right].
\end{eqnarray*}
\end{widetext}

Under the assumptions that the following three conditions are
satisfied
\begin{equation}\label{conditions}
  \beta \gg 1, \ \ \ \ \ \frac{\gamma}{|{\bf u}_r|} \ll 1, \ \ \ \ \
  \frac{1}{|{\bf u}_r|\beta \gamma} \ll 1,
\end{equation}
where ${\bf u}=(\K_1+\K_2)/|\K_1+\K_2|$ and ${\bf u}_r = (\K_{1,r}+\K_{2,r})/|\K_1+\K_2|$ refers to the radial
component of ${\bf u}$.

We show in appendix~\ref{app_G1} that the atomic density is given by
\begin{eqnarray}\label{density}
  G^{(1)}(\K,\K) =\frac{\alpha^2\beta \gamma^3}{32\sqrt{2\pi}|{\bf u}_r|}\exp \left[ - \frac{2\beta^2
      \gamma^2(k-1)^2 }{|{\bf u}_r|^2 }\right],
\end{eqnarray}

and the first order correlation function by

\begin{eqnarray}
  &&G^{(1)}(\K_1,\K_2) =\frac{\alpha^2\beta \gamma^3}{32\sqrt{2\pi}|{\bf u}_r|}\exp\left[-\frac{\gamma^2\beta^2}{8}
    \Delta \K_r^2 - \frac{\beta^2}{8} \Delta k_z^2 \right]   \nonumber\\
  &&  \times \exp \left( - \frac{\beta^2}{8} ({\bf u}\Delta \K)^2 \right)\left(1 - \mbox{erf}\left(
  \frac{i\beta {\bf u}\Delta \K}{2\sqrt{2}} \right) \right)  \nonumber\\
  && \times \exp\left( - \frac{2\beta^2 \gamma^2 \Delta K^2}{{\bf u}_r^2}   \right).\label{gfin}
\end{eqnarray}
We have introduced $\frac{|\K_1 + \K_2|}{2} = 1 +\Delta K$, $\Delta
\K = \K_1 - \K_2$ and assumed $|\Delta\K|$ is small.\\

The conditions (\ref{conditions}) are fulfilled in the experiment of Ref~\cite{perrin:07}
because the region $u_r\sim 0$ corresponds to the location of the two condensates and has
been excluded from the analysis. The density of the scattered particles is peaked around $k=1$
with a width of $\frac{|{\bf u}_r|}{\beta \gamma} \ll 1$. 
We thus expect an anisotropic halo thickness, but the anisotropy is only strong around $u_r \sim 0$,
a direction which was inaccessible in the experiment of Ref.~\cite{perrin:07}

As in the case of the anomalous density $M$, we can decompose $G^{(1)}(\K_1,\K_2)$ 
into factors expressing momentum conservation (1st line of Eq. 11) and energy conservation (2nd line of Eq. 11).
We find that the widths of the momentum contribution are $\sqrt{2}$ larger than the corresponding ones for
$M(\K_1,\K_2)$~\cite{zin:06, molmer:07}.  As discussed
in Refs. [15] and [16], the $\sqrt 2$ is due to the assumption of a Gaussian
density profile. The energy contribution happens to be much more constraining than for $M(\K_1,\K_2)$
because of the term ${\bf u}\Delta \K$. If ${\bf u}\Delta \K= 0$, meaning $k_1=k_2$,
the width of $G^{(1)}(\K_1,\K_2)$ is given by the momentum contribution. 
But, if ${\bf u}\Delta \K\neq 0$ and for instance if ${\bf u}$ is parallel to $\Delta \K$, 
its width is $\propto 1/\beta$ even in the radial plane, in contradiction with
the simple model developed in Ref.\cite{perrin:07}. 

\subsection{Applicability of perturbation theory}
\label{sec2_ap}
Perturbation theory is valid provided the scattering 
of atoms is spontaneous. 
When bosonic
enhancement comes in to play, the perturbative approach fails.
Here we give a simple estimate for parameters such as the number
of scattered atoms and the dimensionless parameter $\beta$ for
which the perturbation is small and the above first order results
can be used.


A coherence volume can be attributed to each scattered atom. 
It is a volume in momentum space in which the atom is first-order
coherent. In other words, if we choose a scattered atom with
momentum $\K$, the volume set by all the wave-vectors $\K_1$ for
which $|G^{(1)}(\K,\K_1)|$ is not negligible is the coherence volume.
If two bosons scatter in such a way  that their
coherence volumes overlap, their joint detection amplitude is enhanced
by an interference effect.
In other words, scattering into an already occupied mode is stimulated. 
The function $G^{(1)}$ permits an estimate of both the number of scattered
 atoms and their associated coherence volumes. If the number of scattered atoms 
 is small, coherence volumes are unlikely to overlap, and stimulated scattering is negligible.
 In this situation we expect our perturbative solution to be valid. 

The above argument was used in the case of the collision of two
spherically symmetric ($\gamma=1$) Gaussian wave-packets
\cite{zin:06} and, in comparisons with numerical solutions of the
equation for the field $\hat\delta$ proved to be correct. Here we
apply an analogous reasoning for the case $\gamma\neq1$.
A conservative estimate for the maximum number of scattered atoms for which 
the perturbative approach applies is $N_{\rm crit}= V/V_c$, 
where $V$ is a lower bound on the $k$-space volume into which atoms 
are scattered, and $V_c$ is an upper bound on the coherence volume of an individual atom. 

In the comparison with the experiment (section~\ref{sec_exp})  we analyze a 
$k$-space volume
$\Omega$ which excludes angles $\theta$ smaller than $\pi/4$.
From Eq.(\ref{density}) one sees that the density of scattered atoms is peaked
around $k=1$ with an rms width of $\sin\theta/\gamma\beta$. In the
volume $\Omega$, the minimum rms width of the shell is $(\gamma\beta\sqrt2)^{-1}$. 
Taking twice
this minimum rms as the thickness of the shell, we find a lower limit
on the volume of $V > 4\pi/\gamma\beta$.



The analysis of Eq. (\ref{gfin}) shows that $V_c$ reaches its
maximum in $\Omega$ for $\theta\simeq\pi/4$ (or
$\theta\simeq3\pi/4$, but due to symmetry we will focus on one of
these values).
If we set $\theta=\pi/4+\delta\theta$,
$\varphi=\delta\varphi$ and $k_1=k_2=1$ we find
\begin{eqnarray*}
  G^{(1)}(\theta,\ \varphi)\propto
  \exp\left(-\frac{\beta^2(\delta\theta)^2}{16}-\frac{\beta^2\gamma^2(\delta\varphi)^2}{16}\right).\nonumber
\end{eqnarray*}
This gives an angular area of coherence approximately equal to
8$\pi/\gamma\beta^2$. Now we need to find the coherence width in the
radial direction. Setting $\K_1=(1+\delta k/2)\K/k$ and $\K_2=(1-\delta k/2)\K/k$ we get:

\begin{eqnarray*}
  G^{(1)}(\delta k)\propto
  \exp\left(-\frac{\beta^2\delta k^2}{8}\right).\nonumber
\end{eqnarray*}
The limit on the coherence volume is therefore:
$V_c < 64\pi/3\gamma\beta^3$.

Combining the estimates of $V$ and $V_c$, we find that critical number of
atoms is given by $N_{\mathrm{crit}}=\frac{3\beta^2}{16}$.
For $\beta=227$ we get $N_{\mathrm{crit}}\approx 10^4$. 
In the experimental realization, the number of atoms detected in $\Omega$
varied from 30 to 300. Assuming 10\% detection efficiency this
gives a maximum of 3000 scattered atoms.
Thus the experiment should be in the perturbative regime. 
A similar argument is given in Ref.~\cite{perrin:08}  leading to a similar value of  $N_{{\rm crit}}$. 


\section{Comparison with experiment}\label{sec_exp}

The formulae (\ref{m_final}) and (\ref{gfin}) cannot be directly compared with
experimental data. This is due to an extra step which is made
during the measurements:
the joint probabilities measured in experiment are averaged over a 
region of interest $\Omega$ which excludes the unscattered condensates. 
We approximate $\Omega$ by
$\theta\in[\frac{\pi}{4},\frac{3\pi}{4}]$, 
$\varphi\in[0,2\pi]$ (where ${\bf u}=(\sin\theta\cos\varphi,\sin\theta\sin\varphi,\cos\theta$)).

In case of
local momentum correlations, the normalization procedure is done by choosing $\K_1$ and
$\K_2$  almost equal:
$\K_1-\K_2=\delta\K$, where $\delta\K$ is small. So we set $\K_1=\K+\delta\K/2$ and
$\K_2=\K-\delta\K/2$. The averaging corresponds to calculation of an integral
\begin{equation}
  \langle |G^{(1)}(\delta\K)|^2\rangle=\int_\Omega d\K\ |G^{(1)}(\K_1,\K_2)|^2.\label{ave}
\end{equation}
Then, this function is normalized by
\begin{equation}
  \int_\Omega d\K\ G^{(1)}(\K_1,\K_1)\cdot G^{(1)}(\K_2,\K_2).\label{norma}
\end{equation}
Let's denote the resulting normalized function by $\langle |g^{(1)}(\delta \K)|^2\rangle$. As the
anomalous density vanishes for local correlations, Eq.(\ref{korelacja}) gives
\begin{eqnarray*}
  g^{(2)}(\delta \K)=1+\langle |g^{(1)}(\delta \K)|^2\rangle.\nonumber
\end{eqnarray*}
For $\delta\K=0$ we get $g^{(2)}(0)=2$ \cite{molmer:07}.

In case of
back-to-back momentum correlations, in analogy we have $\K_1$ and
$\K_2$  almost opposite:
$\K_1+\K_2=\delta\K$. We set $\K_1=\K+\delta\K/2$ and
$\K_2=-\K+\delta\K/2$. Once again, the averaging corresponds to
\begin{eqnarray*}
  \langle |M(\delta\K)|^2\rangle=\int_\Omega d\K\ |M(\K_1,\K_2)|^2.\nonumber
\end{eqnarray*}
After normalization by function (\ref{norma}) we obtain  $\langle |m(\delta \K)|^2\rangle$. For the
opposite momentum correlations, $G^{(1)}$ vanishes, thus
\begin{eqnarray*}
  g^{(2)}(\delta \K)=1+\langle |m(\delta \K)|^2\rangle.\nonumber
\end{eqnarray*}
Let us now calculate the normalization function from (\ref{norma}), as it is common for both local- and opposite-
momentum correlations. From Eq.(\ref{gfin}) we have
\begin{eqnarray*}
  G^{(1)}(\K_{1,2})=\frac{\alpha^2\beta \gamma^3
    \sqrt{\pi}}{32\pi\sqrt{2}|{\bf u}_{1,2r}|} \exp \left[ - \frac{\beta^2
      \gamma^2(k_{1,2}^2-1)^2 }{2|{\bf u}_{1,2r}|^2 }\right].
\end{eqnarray*}
Now, in spherical coordinates, $|{\bf u}_{1,2r}|=|\sin\theta_{1,2}|$, where $\theta_{1,2}$
is an angle between vector $\K_{1,2}$ and axis $z$.
Since $\frac{1}{2}\delta\K$ is much smaller than $\K$, 
we can approximate $\sin\theta_{1,2}\simeq\sin\theta$,
where $\theta$ is an angle between vector $\K$ and axis $z$ 
and drop higher order terms in $\delta\K$ in the exponentials. 
We end up with the approximate expression
\begin{eqnarray*}
  &&\int_\Omega d\K\ G^{(1)}(\K_{1},\K_{1})\cdot G^{(1)}(\K_{2},\K_{2})
  \simeq\frac{\alpha^4\beta^2 \gamma^6}{2^{11}\pi\sin^2\theta}\\
  &&\times\int_\Omega d\K\ \exp\left[-\frac{\beta^2\gamma^2(k^2-1)^2 }{\sin^2\theta }
    - \frac{\beta^2\gamma^2(\K\cdot\delta\K)^2}{\sin^2\theta}\right].\nonumber
\end{eqnarray*}
If $\delta\K=\delta k\cdot {\bf e}_x$, $\K\cdot\delta\K=k\delta k\sin\theta\cos\phi$ and if 
$\delta\K=\delta k\cdot{\bf e}_z$, $\K\cdot\delta\K=k\delta k\cos\theta$. 
The resulting integrals are calculated numerically.


\subsection{Back to back momentum correlations}

As discussed above, we set 
$\K_1=\K+\delta\K/2$ and
$\K_2=-\K+\delta\K/2$. Using Eq.(\ref{m_final})
\begin{eqnarray*}
  &&|M(\K_1,\K_2)|^2=\frac{\alpha^2\beta^4\gamma^4}{256\pi^2}  \exp\left(
  -\frac{\gamma^2\beta^2}{2}\delta k_{r}^2 -\frac{\beta^2}{2}\delta k_{z}^2- \frac{ \Delta^2}{4}\right)\nonumber\\
  &&
  \left( 1 + \mbox{erfi}^2 \left( \frac{\Delta}{2} \right)  \right).
\end{eqnarray*}
Here, $\Delta=\beta(1-k^2-\frac{\delta k^2}{4})$. The
averaging over $\Omega$ is equivalent to
\begin{eqnarray*}
  &&\langle |M(\delta \K)|^2\rangle=\int_\Omega d\K\ |M(\K_1,\K_2)|^2=\frac{\alpha^2\beta^4\gamma^4}{256\pi^2}
      e^{-\frac{\gamma^2\beta^2}{2}\delta k_{r}^2 }\nonumber\\
  &&\times e^{-\frac{\beta^2}{2}\delta k_{z}^2}\int_\Omega d\K\ e^{-\frac{ \Delta^2}{4}}
  \left( 1 + \mbox{erfi}^2 \left( \frac{\Delta}{2} \right)  \right).
\end{eqnarray*}

\begin{figure}[htb]
  \centering
  \includegraphics[scale=0.32, angle=0]{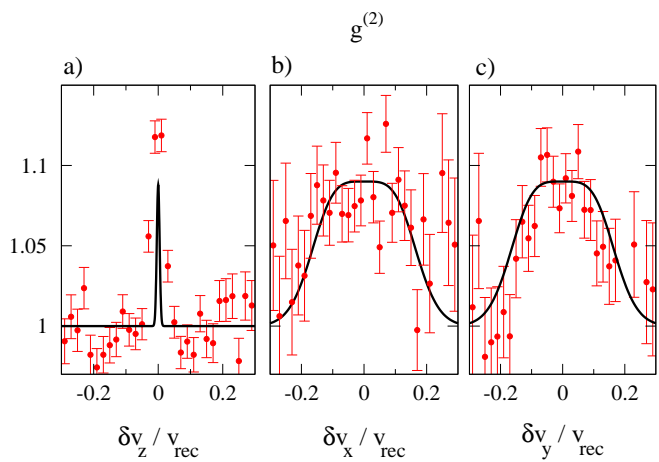}
  \caption{(color online): Normalized opposite momentum correlations calculated in perturbative regime as compared with experimental
    data. Three plots correspond to three different directions. Here, $\delta v_i=\hbar/m\cdot \delta k_i$ and
    $v_{rec}=\hbar/m\cdot Q$.} \label{back}
\end{figure}

Numerical evaluation of this integral (for parameters $\beta$ and
$\gamma$ as defined above) shows that the averaged anomalous density can be well-approximated by
\begin{eqnarray*}
  \langle |M(\delta \K)|^2\rangle\propto\exp\left(-\frac{\gamma^2\beta^2}{2}\delta k_{r}^2
  -\frac{\beta^2}{2}\delta k_{z}^2\right).\nonumber
\end{eqnarray*}
As we see, the width of $\langle |M(\delta \K)|^2\rangle$ is primarily determined by the momentum conservation
constraint, but the analysis shows that energy conservation plays a role, 
decreasing the predicted width in the $xy$-plane by of order 10\%.
We normalize the second order correlation function by (\ref{norma}) and introduce 
an empirical parameter $\eta_\mathrm{bb}$ to account for the fact that in the
experimental data plots, the correlation functions are projections, and the fact that their heights
were smaller than expected. 
%
%
We find
\begin{eqnarray*}
  g^{(2)}(\delta \K)=1+\eta_\mathrm{bb}\langle |m(\delta \K)|^2\rangle.\nonumber
\end{eqnarray*}
This function is plotted in Fig.~\ref{back}, using the value  $\eta_\mathrm{bb}=0.032$. 
We find good agreement with the experimental data in the $x$- and $y$- directions.
In the $z$-direction, the width of the experimental peak is dominated by the detector 
resolution which is larger than the calculated width. 

\subsection{Local momentum correlations}

For the collinear correlation function we choose $\K_1=\K+\delta\K/2$ and $\K_2=\K-\delta\K/2$.
Using Eq.(\ref{gfin}) and definition from Eq.(\ref{ave}) we have
\begin{eqnarray*}\nonumber
  &&\langle |G^{(1)}(\delta\K)|^2\rangle=\int_\Omega d\K\
  \frac{\alpha^4\beta^2 \gamma^6}{2^{11}\pi|{\bf u}_r|^{2}}
  \exp \left( - \frac{\beta^2}{4} ({\bf u}\delta \K)^2 \right)\\
  &&\exp\left[-\frac{\gamma^2\beta^2}{4} \delta \K_r^2 - \frac{\beta^2}{4} \delta k_z^2 \right]
  \left[1 + \mbox{erfi}^2\left(\frac{\beta}{2\sqrt{2}}{\bf u}\delta \K \right) \right]\nonumber  \\
  && \times \exp\left( - \frac{4\beta^2 \gamma^2  (k-1)^2}{{\bf u}_r^2}   \right).
\end{eqnarray*}
Let us now consider two separate cases.

Let's set $\delta\K=\delta k_x{\bf e}_x$. Then, ${\bf
u}\delta\K=\delta k\sin\theta\cos\varphi$.
Integration over the
region $\Omega$ consists of an angular and a radial integral. The
radial one is
\begin{eqnarray*}
  I_r=\int_0^\infty k^2dk\times\exp\left(-\frac{4\beta^2 \gamma^2  (k-1)^2}{{\bf u}_r^2}\right).\nonumber
\end{eqnarray*}
The width of this Gaussian function is so small, that we can set
$k^2dk\sim dk$. Setting $k=1+dk$ and extending the lower limit of
the integral to $-\infty$ gives $I_r\propto|{\bf u}_r|$.
Thus
\begin{eqnarray*}
  &&\langle|G^{(1)}(\delta k_x)|^2\rangle \propto
  e^{-\frac{\gamma^2\beta^2}{4}\delta k_x^2}\int_0^{2\pi}d\varphi\int_{\pi/4}^{3/4\pi}d\theta\\
  &&  \exp\left( - \frac{\beta^2}{4}\delta k_x^2\cdot u^2(\theta,\varphi)\right)\left[1 + \mbox{erfi}^2\left(
      \frac{\beta\cdot\delta k_x}{2\sqrt{2}} u(\theta,\varphi) \right) \right],
\end{eqnarray*}
where $u(\theta,\varphi)=\sin\theta\cos\varphi$.
This integral is calculated numerically and we obtain
\begin{eqnarray*}
  \langle g^{(2)}(\delta k_x)\rangle=1+\langle|g^{(1)}(\delta k_x)|^2\rangle.\nonumber
\end{eqnarray*}
%
The result is again rescaled by the parameter $\eta_{\mathrm cl}$ although it needs not to be 
identical to the back to back case:
\begin{eqnarray*}
  \langle g^{(2)}(\delta k_x)\rangle=1+\eta_{\mathrm cl}\langle|g^{(1)}(\delta k_x)|^2\rangle.\nonumber
\end{eqnarray*}
As $\langle|g^{(1)}(0)|^2\rangle=1$, we deduce the value of $\eta_\mathrm{cl}=0.05$.

Now we set $\delta\K=\delta k_z{\bf e}_z$, and therefore ${\bf
u}\delta\K=\delta k_z\cos\theta$.
The radial integral is the same as in the previous case and we find
\begin{eqnarray*}
  &&\langle|G^{(1)}(\delta k_z)|^2\rangle \propto
  \exp\left[-\frac{\beta^2}{4} (\delta k_z)^2\right]\int_{\pi/4}^{3/4\pi}d\theta\\
  &&  \exp \left( - \frac{\beta^2}{4}(\delta k_z)^2\cdot \cos^2\theta \right)\left[1 + \mbox{erfi}^2\left(
      \frac{\beta\cdot\delta k_z}{2\sqrt{2}} \cos\theta \right) \right].
\end{eqnarray*}
Numerically we find:
\begin{eqnarray*}
  \langle g^{(2)}(\delta k_z)\rangle=1+\eta_\mathrm{cl}\langle|g^{(1)}(\delta k_z)|^2\rangle.\nonumber
\end{eqnarray*}
We find that chosing $\eta_\mathrm{cl}=0.05$ makes the observed heights match.


\begin{figure}[htb]
  \centering
  \includegraphics[scale=0.32, angle=0]{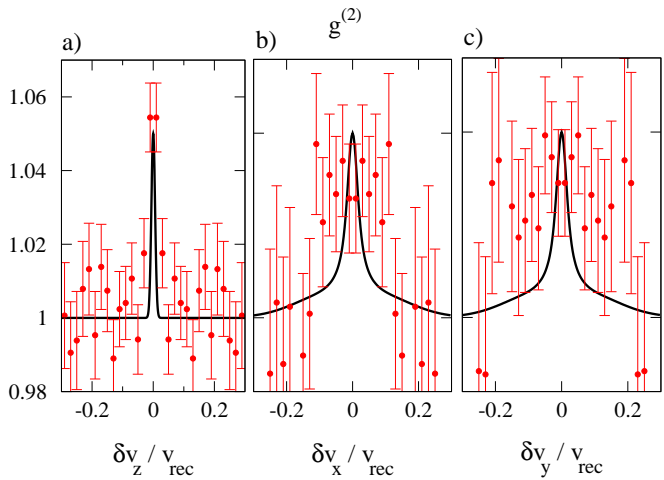}
  \caption{(color online): 
    Normalized collinear correlations calculated in perturbative regime as compared with experimental
    data. Three plots correspond to three different directions. Due to cylindrical symmetry of the colliding
    condensates, theoretical results preserve this symmetry. Here, $\delta v_i=\hbar/m\cdot \delta k_i$ and
    $v_{rec}=\hbar/m\cdot Q$.}\label{cl}
\end{figure}

Once again, because of the detector resolution, 
we find that the calculated peak is much narrower that the observed one in
the $z$-direction. What is more surprising is that the widths of the correlation
functions in the $x$- and $y$-directions are also narrower than those in the experiment. 
As can be seen from the discussion following Eq.(\ref{gfin}), the peak width along the 
direction of the outgoing atoms is strongly constrained by the energy conservation  
requirement. 
This means that for scattering far from the $z$-axis ($\theta$ large), the 
$x$- and $y$-components of the correlation function are narrower than 
they would be taking momentum conservation alone into account.
This result contradicts the simple reasoning of Ref.~\cite{perrin:07}. 
In the next section we speculate about why neither the experiment nor the
positive P simulation results reproduce the above width for the correlation function.



\section{Conclusions}

The perturbative result we have presented here, while rather complex, 
has the virtue that the results are analytic and permit the identification the physical processes involved in the pair formation process. 
In particular the roles of energy and momentum conservation are clearly identified. 
Our results for the back to back correlation are in good agreement with the experiment. 
On the other hand the collinear correlation function, as shown in Fig.~\ref{cl}, 
is in apparent contradiction with both the experiment and with the calculation of Ref.~\cite{perrin:08}. 
The perturbative correlation function given in this work is narrower.  
This discrepancy clearly needs more attention, both theoretical and experimental, but we wish to make some comments about possible causes. 
First, as discussed in Ref.~\cite{perrin:08}, the calculations using the 
positive P representation are not able to simulate the entire duration of the collision; 
indeed only about 20\% 
of the collision time can be simulated. 
Thus, energy conservation is not as strictly enforced leading to additional broadening in the calculations of 
Ref.~\cite{perrin:08}. Although this effect was discussed in that reference, 
the problem requires further scrutiny, 
it is not entirely clear to us which widths are most affected by a short collision time. 
Second, the experimental observations are also subject to effects not treated here. 
It was briefly mentioned in Ref.~\cite{perrin:07} 
that the mean field interaction between the escaping atoms and the 
remaining condensates may not be negligible. 
It is therefore important to undertake an analysis of their effect on the correlation functions. 
Finally, an important simplification in the present treatment is the
assumption that the condensates do not expand during the collision.
This assumption seems reasonable because most of the atom collisions take place 
before the clouds have had time to expand.
Still, a quantitative estimate of the influence of the condensate
expansion is another avenue for future analysis.

 Clarifying these questions may have ramifications beyond atom optics.
Conceptually similar experiments involving collisions between heavy
ions have also uncovered discrepancies between observations and simple
models \cite{lisa:05, wong:07}, the so-called ``HBT puzzle". 
We hope that the work presented here
will continue to stimulate careful thought about the four wave mixing
process of matter waves.

\section{Acknowledgements}

We acknowledge the support of the CIGMA project of the Eurocores program of ESF, 
the SCALA project of the EU and the Institut Francilien pour la Recherche en Atomes Froids. 
P Z. and J. Ch. acknowledge the support of Polish Government scientific
grant (2007-2010).
\appendix

\section{Determination of $\alpha$}\label{app_alpha}

When we introduced $\alpha$, it was simply defined in terms of the
number of atoms, the condensate size and the scattering length. Here
we give a complementary estimate of $\alpha$ which provides a
consistency check. The result essentially shows that our treatment is able to
predict, to within experimental uncertainties, the number of scattered
atoms. We start from Eq.(\ref{density}).
The integration of this
equation over $\Omega$ gives the number of scattered atoms to
first order. This result, being a function of $\alpha$, can be
compared with the number of scattered atoms in the experiment.
Knowing this number, we can evaluate $\alpha$. First, using Eq.(\ref{density}), 
the number of scattered atoms in $\Omega$ is
given by
\begin{eqnarray*}
  \mathcal{N}_\Omega = \frac{\alpha^2\beta \gamma^3}{32\sqrt{2\pi}|{\bf u}_r|}\int_\Omega d\K\exp \left[ -
    \frac{\beta^2 \gamma^2(k^2-1)^2 }{2|{\bf u}_r|^2 }\right].\nonumber
\end{eqnarray*}
Let us focus for a moment on the radial part of the above
integral,
\begin{eqnarray*}
  I_{\mathrm{rad}}=\frac{\alpha^2\beta \gamma^3}{32\sqrt{2\pi}|{\bf u}_r|}
  \int_0^\infty k^2dk\exp
  \left[ - \frac{\beta^2 \gamma^2(k^2-1)^2 }{2|{\bf u}_r|^2
    }\right].\nonumber
\end{eqnarray*}
First, as the integrand is strongly peaked around $k=1$, the
measured volume can be dropped, i.e. $k^2\sim1$. Then, introducing
$k=1+\delta k$ and assuming $\delta k$ is small we get
\begin{eqnarray*}
  I_{\mathrm{rad}}\simeq\frac{\alpha^2\beta \gamma^3}{32\sqrt{2\pi}|{\bf u}_r|}\int_{-1}^\infty d(\delta k)
  \exp \left[-\frac{\beta^2 \gamma^2(\delta k)^2 }{2|{\bf u}_r|^2
    }\right].\nonumber
\end{eqnarray*}
The lower limit can be extended to $-\infty$, giving
\begin{eqnarray*}
  I_{\mathrm{rad}}\simeq\frac{\alpha^2\beta \gamma^3}{32\sqrt{2\pi}|{\bf u}_r|}\int_{-\infty}^\infty
  d(\delta k) \exp \left[-\frac{\beta^2 \gamma^2(\delta k)^2
    }{2|{\bf u}_r|^2 }\right]=\frac{\alpha^2\gamma^2}{64}.\nonumber
\end{eqnarray*}
Integration over the angular variables gives a factor of $2\sqrt
2\pi$ and
\begin{eqnarray*}\nonumber
  \mathcal{N}_\Omega =\frac{\pi\sqrt2}{32}\alpha^2\gamma^2.
\end{eqnarray*}
From the experimental data we know that the number of
scattered atoms varies from 300 to 3000. For $N_\Omega=300$ we get
$\alpha=930$ and for $N_\Omega=3000$ we get $\alpha=2940$. Thus
the value of $\alpha=1053$ calculated from the model of colliding
Gaussians lies somewhere in between. This result,
confirms that the choice of parameters such as $\sigma_r$ and
$\sigma_z$ are reasonable.

\section{First order correlation function: $\K_1\simeq \K_2$ correlations}\label{app_G1}
To first order the $G^{(1)}$ function is,
\begin{widetext}
\begin{eqnarray*}
  && G^{(1)}(\K_1,\K_2)=|\mathcal{A}|^2\int_0^\infty \mbox{d} \tau\int_0^\infty \mbox{d}\tau' \,
  \exp\left[-\tau^2-\tau'^2+i\beta(\tau-\tau')\right]\int d\K \,
  \exp\left[-\frac{\gamma^2\beta^2}{4}\left((\K_{1,r}+\K_r)^2+(\K_{2,r}+\K_r)^2\right)\right]\nonumber\\
  &&\times\exp\left[-\frac{\beta^2}{4}\left((k_{1,z}+k_z)^2+(k_{2,z}+k_z)^2\right)\right]
  \exp\left[i\frac{\beta}{2}(k^2+k_2^2)\tau'-i\frac{\beta}{2}(k^2+k_1^2)\tau\right].
\end{eqnarray*}
\end{widetext}

Thus, in contrast to the anomalous density, we must perform a
two-fold time as well as a three-dimensional space integral. The
space integral can be evaluated analytically. Then, introducing $x
= \frac{\tau + \tau'}{\sqrt{2}}$ and  $ y = \frac{\tau -
\tau'}{\sqrt{2}}$ the first order correlation function is
\begin{widetext}
  \begin{eqnarray*}\nonumber
    &&G^{(1)}(\K_1,\K_2)  =\frac{\alpha^2\beta
      \gamma^2}{16\pi^{3/2}\sqrt{2}}
    \exp\left[-\frac{\gamma^2\beta^2}{8} |\K_{1,r} - \K_{2,r}|^2 - \frac{\beta^2}{8} |k_{1,z} - k_{2,z}|^2 \right]
    \int_0^\infty \mbox{d} x \int_{-x}^x \mbox{d} y \, \exp\left[-x^2+ i\frac{\beta}{2\sqrt{2}}x (k_2^2-k_1^2) \right]
    \\ \nonumber
    && \times\exp\left[ -y^2 \left( 1 + \frac{{\bf u}_r^2}{\gamma^2} + {\bf u}_z^2 \right)+i\beta \sqrt{2} y \left( 1-
      \frac{k_1^2+k_2^2}{4} - \frac{(\K_1+\K_2)^2}{8} \right) \right]
    \exp\left[ i\sqrt{2} y \frac{(\K_{1,r}+\K_{2,r})^2}{4} \frac{y^2}{\gamma^2 (\beta
        \gamma^2)}\frac{1}{1+ 2y^2/(\beta \gamma^2)^2} \right]
    \\ \nonumber
    && \times
    \exp\left[  \frac{y^2}{\gamma^2} \left({\bf u}_r^2 -\frac{(\K_{1,r}+\K_{2,r})^2}{4(1+ 2y^2/(\beta
        \gamma^2)^2)}  \right)  \right]
    \exp\left[ i\sqrt{2} y\frac{(k_{1,z}+k_{2,z})^2}{4} \frac{y^2}{\beta} \frac{1}{1+
        2y^2/\beta^2} \right]
    \exp\left[  y^2 \left({\bf u}_z^2 - \frac{(\K_{1,r}+\K_{2,r})^2}{4(1+ 2y^2/\beta^2)}  \right)  \right]
    \\
    && \times
    \frac{1}{1 + i\sqrt{2} y/(\beta \gamma^2) } \frac{1}{\sqrt{1 + i \sqrt{2} y/\beta  }}
  \end{eqnarray*}
\end{widetext}
where ${\bf u}={\bf u}_r+{\bf u}_z$ is a vector of unit length and
direction $\K_1 + \K_2$. As the scattering of atoms conserves
energy and momentum, we expect that the density of atoms should be
centered around $|\K|=1$ (which corresponds to $|\K|=Q$ in
physical units). Moreover, from the factor $\exp\left[ -y^2 \left(
1 + \frac{{\bf u}_r^2}{\gamma^2} + {\bf u}_z^2 \right)\right]$, we
deduce that the characteristic width of variable $y$ is $1/\sqrt{1
+ \frac{{\bf u}_r^2}{\gamma^2}+ {\bf u}_z^2}$.

Using the second of conditions (\ref{conditions}) we have
\begin{eqnarray*}
  \exp\left[ -y^2 \left( 1 + \frac{{\bf u}_r^2}{\gamma^2} + {\bf
      u}_z^2 \right)\right]\simeq \exp\left[ -y^2\frac{{\bf
        u}_r^2}{\gamma^2}\right].\nonumber
\end{eqnarray*}
Since the characteristic range of $y$ is $\gamma/|{\bf u}_r|$,
 all the terms proportional to $y/\beta$ and $y/\beta\gamma^2$ can be dropped. This gives
\begin{widetext}
\begin{eqnarray}
  &&G^{(1)}(\K_1,\K_2) =\frac{\alpha^2\beta \gamma^2}{16\pi^{3/2}\sqrt{2}}
  \exp\left[-\frac{\gamma^2\beta^2}{8} |\K_{1,r} - \K_{2,r}|^2 - \frac{\beta^2}{8} |k_{1,z} - k_{2,z}|^2 \right]
  \int_0^\infty \mbox{d} x \int_{-x}^x \mbox{d} y \, \exp\left[-x^2+ i\frac{\beta}{2\sqrt{2}}
    x (k_2^2-k_1^2) \right]  \nonumber
\\
  &&\times\exp\left[ -\frac{y^2}{\gamma^2}{\bf u}_r^2 +i\beta \sqrt{2} y \left( 1- \frac{k_1^2+k_2^2}{4} -
    \frac{(\K_1+\K_2)^2}{8} \right) +
  \frac{y^2}{\gamma^2} \left({\bf u}_r^2 -\frac{(\K_{1,r}+\K_{2,r})^2}{4}  \right)
    \right].\label{g1_big2}
\end{eqnarray}
\end{widetext}
Now, by letting $\K_1=\K_2=\K$ in Eq.(\ref{g1_big2}) let us focus
on the momentum density of scattered atoms,
\begin{eqnarray*}\nonumber
  &&G^{(1)}(\K,\K) =\frac{\alpha^2\beta
    \gamma^2}{16\pi^{3/2}\sqrt{2}} \int_0^\infty \mbox{d} x \,
  \exp\left[-x^2 \right]
  \\
  &&\nonumber \int_{-x}^x \mbox{d} y \,  \exp\left[
    -\frac{y^2}{\gamma^2}{\bf u}_r^2+i\beta \sqrt{2} y \left( 1- k^2
    \right) \right]\\ \nonumber && \times \exp\left[
    \frac{y^2}{\gamma^2} {\bf u}_r^2(1-k^2)  \right]
\end{eqnarray*}
From the above we deduce that the characteristic width of $x$ is
$1$ which is much larger that the characteristic width of $y$.
This allows another approximation -- the limits of $y$
integral can be expanded up from $-\infty $ to $\infty$. The
variables $y$ and $x$ effectively decouple, giving
\begin{eqnarray*}\nonumber
  &&G^{(1)}(\K,\K) =\frac{\alpha^2\beta \gamma^2}{32\pi\sqrt{2}}
  \int_{-\infty}^\infty \exp\left[ -\frac{y^2}{\gamma^2}{\bf
      u}_r^2+i\beta \sqrt{2} y \left( 1- k^2 \right) \right]
  \\\nonumber
&& \times \exp\left[  \frac{y^2}{\gamma^2} {\bf u}_r^2 (1-k^2)
    \right]\mbox{d} y.
\end{eqnarray*}

After integration over $y$ and with $k\sim 1$, one obtains,
\begin{eqnarray*}
  G^{(1)}(\K,\K) =\frac{\alpha^2\beta \gamma^3}{32\sqrt{2\pi}|{\bf u}_r|}\exp \left[ - \frac{2\beta^2
      \gamma^2(k-1)^2 }{|{\bf u}_r|^2 }\right].
\end{eqnarray*}

Equation (\ref{g1_big2}) can be rewritten in the form
\begin{widetext}
  \begin{eqnarray*}\nonumber
    &&G^{(1)}(\K_1,\K_2) =\frac{\alpha^2\beta \gamma^2}{16\pi^{3/2}\sqrt{2}}
    \exp\left[-\frac{\gamma^2\beta^2}{8} |\K_{1,r} - \K_{2,r}|^2 - \frac{\beta^2}{8} |k_{1,z} - k_{2,z}|^2 \right]
    \int_0^\infty \mbox{d} x \int_{-x}^x \mbox{d} y \, \exp\left[-x^2+ i\frac{\beta}{2\sqrt{2}}x (k_2^2-k_1^2) \right]
    \\ \nonumber
    && \times\exp\left[ -\frac{y^2}{\gamma^2}{\bf u}_r^2 +i\beta \sqrt{2} y \left( 1- \frac{k_1^2+k_2^2}{4} -
      \frac{(\K_1+\K_2)^2}{8} \right)
      +\frac{y^2}{\gamma^2} \left({\bf u}_r^2 -\frac{(\K_{1,r}+\K_{2,r})^2}{4}  \right)\right].
  \end{eqnarray*}
\end{widetext}
Introducing $\frac{|\K_1 + \K_2|}{2} = 1 +\Delta K$ and  $\Delta
\K = \K_1 - \K_2$, where $|\Delta\K|$ is small we obtain
\begin{eqnarray*}
  &&G^{(1)}(\K_1,\K_2) =\frac{\alpha^2\beta \gamma^3}{32\sqrt{2\pi}|{\bf u}_r|}\exp\left[-\frac{\gamma^2\beta^2}{8}
    \Delta \K_r^2 - \frac{\beta^2}{8} \Delta k_z^2 \right]   \nonumber\\
  &&  \times \exp \left( - \frac{\beta^2}{8} ({\bf u}\Delta \K)^2  - \frac{2\beta^2 \gamma^2 \Delta K^2}{{\bf u}_r^2}   
  \right)\left(1 - \mbox{erf}\left(
  \frac{i\beta {\bf u}\Delta \K}{2\sqrt{2}} \right) \right).
\end{eqnarray*}

\bibliographystyle{apsrev}

\begin{thebibliography}{28}
\expandafter\ifx\csname natexlab\endcsname\relax\def\natexlab#1{#1}\fi
\expandafter\ifx\csname bibnamefont\endcsname\relax
  \def\bibnamefont#1{#1}\fi
\expandafter\ifx\csname bibfnamefont\endcsname\relax
  \def\bibfnamefont#1{#1}\fi
\expandafter\ifx\csname citenamefont\endcsname\relax
  \def\citenamefont#1{#1}\fi
\expandafter\ifx\csname url\endcsname\relax
  \def\url#1{\texttt{#1}}\fi
\expandafter\ifx\csname urlprefix\endcsname\relax\def\urlprefix{URL }\fi
\providecommand{\bibinfo}[2]{#2}
\providecommand{\eprint}[2][]{\url{#2}}

\bibitem[{\citenamefont{Burnham and Weinberg}(1970)}]{burnham:70}
\bibinfo{author}{\bibfnamefont{D.~C.} \bibnamefont{Burnham}} \bibnamefont{and}
  \bibinfo{author}{\bibfnamefont{D.~L.} \bibnamefont{Weinberg}},
  \bibinfo{journal}{Phys. Rev. Lett.} \textbf{\bibinfo{volume}{25}},
  \bibinfo{pages}{84} (\bibinfo{year}{1970}).

\bibitem[{\citenamefont{Greiner et~al.}(2005)\citenamefont{Greiner, Regal,
  Stewart, and Jin}}]{greiner:05}
\bibinfo{author}{\bibfnamefont{M.}~\bibnamefont{Greiner}},
  \bibinfo{author}{\bibfnamefont{C.~A.} \bibnamefont{Regal}},
  \bibinfo{author}{\bibfnamefont{J.~T.} \bibnamefont{Stewart}},
  \bibnamefont{and} \bibinfo{author}{\bibfnamefont{D.~S.} \bibnamefont{Jin}},
  \bibinfo{journal}{Phys. Rev. Lett.} \textbf{\bibinfo{volume}{94}},
  \bibinfo{eid}{110401} (\bibinfo{year}{2005}).

\bibitem[{\citenamefont{Perrin et~al.}(2007)\citenamefont{Perrin, Chang,
  Krachmalnicoff, Schellekens, Boiron, Aspect, and Westbrook}}]{perrin:07}
\bibinfo{author}{\bibfnamefont{A.}~\bibnamefont{Perrin}},
  \bibinfo{author}{\bibfnamefont{H.}~\bibnamefont{Chang}},
  \bibinfo{author}{\bibfnamefont{V.}~\bibnamefont{Krachmalnicoff}},
  \bibinfo{author}{\bibfnamefont{M.}~\bibnamefont{Schellekens}},
  \bibinfo{author}{\bibfnamefont{D.}~\bibnamefont{Boiron}},
  \bibinfo{author}{\bibfnamefont{A.}~\bibnamefont{Aspect}}, \bibnamefont{and}
  \bibinfo{author}{\bibfnamefont{C.~I.} \bibnamefont{Westbrook}},
  \bibinfo{journal}{Phys. Rev. Lett.} \textbf{\bibinfo{volume}{99}},
  \bibinfo{eid}{150405} (\bibinfo{year}{2007}),

\bibitem[{\citenamefont{Duan et~al.}(2000)\citenamefont{Duan, S\o{}rensen,
  Cirac, and Zoller}}]{duan:00}
\bibinfo{author}{\bibfnamefont{L.-M.} \bibnamefont{Duan}},
  \bibinfo{author}{\bibfnamefont{A.}~\bibnamefont{S\o{}rensen}},
  \bibinfo{author}{\bibfnamefont{J.~I.} \bibnamefont{Cirac}}, \bibnamefont{and}
  \bibinfo{author}{\bibfnamefont{P.}~\bibnamefont{Zoller}},
  \bibinfo{journal}{Phys. Rev. Lett.} \textbf{\bibinfo{volume}{85}},
  \bibinfo{pages}{3991} (\bibinfo{year}{2000}).

\bibitem[{\citenamefont{Pu and Meystre}(2000)}]{pu:00}
\bibinfo{author}{\bibfnamefont{H.}~\bibnamefont{Pu}} \bibnamefont{and}
  \bibinfo{author}{\bibfnamefont{P.}~\bibnamefont{Meystre}},
  \bibinfo{journal}{Phys. Rev. Lett.} \textbf{\bibinfo{volume}{85}},
  \bibinfo{pages}{3987} (\bibinfo{year}{2000}).

\bibitem[{\citenamefont{Opatrn\'y and Kurizki}(2001)}]{opatrny:01}
\bibinfo{author}{\bibfnamefont{T.}~\bibnamefont{Opatrn\'y}} \bibnamefont{and}
  \bibinfo{author}{\bibfnamefont{G.}~\bibnamefont{Kurizki}},
  \bibinfo{journal}{Phys. Rev. Lett.} \textbf{\bibinfo{volume}{86}},
  \bibinfo{pages}{3180} (\bibinfo{year}{2001}).

\bibitem[{\citenamefont{Kheruntsyan et~al.}(2005)\citenamefont{Kheruntsyan,
  Olsen, and Drummond}}]{kheruntsyan:05}
\bibinfo{author}{\bibfnamefont{K.~V.} \bibnamefont{Kheruntsyan}},
  \bibinfo{author}{\bibfnamefont{M.~K.} \bibnamefont{Olsen}}, \bibnamefont{and}
  \bibinfo{author}{\bibfnamefont{P.~D.} \bibnamefont{Drummond}},
  \bibinfo{journal}{Phys. Rev. Lett.} \textbf{\bibinfo{volume}{95}},
  \bibinfo{pages}{150405} (\bibinfo{year}{2005}).

\bibitem[{\citenamefont{Vardi and Moore}(2002)}]{vardi:02}
\bibinfo{author}{\bibfnamefont{A.}~\bibnamefont{Vardi}} \bibnamefont{and}
  \bibinfo{author}{\bibfnamefont{M.~G.} \bibnamefont{Moore}},
  \bibinfo{journal}{Phys. Rev. Lett.} \textbf{\bibinfo{volume}{89}},
  \bibinfo{pages}{090403} (\bibinfo{year}{2002}).

\bibitem[{\citenamefont{Bach et~al.}(2002)\citenamefont{Bach, Trippenbach, and
  Rza\ifmmode \mbox{\c{}}\else \c{}\fi{}\ifmmode~\dot{z}\else
  \.{z}\fi{}ewski}}]{bach:02}
\bibinfo{author}{\bibfnamefont{R.}~\bibnamefont{Bach}},
  \bibinfo{author}{\bibfnamefont{M.}~\bibnamefont{Trippenbach}},
  \bibnamefont{and} \bibinfo{author}{\bibfnamefont{K.}~\bibnamefont{Rza\ifmmode
  \mbox{\c{}}\else \c{}\fi{}\ifmmode~\dot{z}\else \.{z}\fi{}ewski}},
  \bibinfo{journal}{Phys. Rev. A} \textbf{\bibinfo{volume}{65}},
  \bibinfo{pages}{063605} (\bibinfo{year}{2002}).

\bibitem[{\citenamefont{Yurovsky}(2002)}]{yurovsky:02}
\bibinfo{author}{\bibfnamefont{V.~A.} \bibnamefont{Yurovsky}},
  \bibinfo{journal}{Phys. Rev. A} \textbf{\bibinfo{volume}{65}},
  \bibinfo{pages}{033605} (\bibinfo{year}{2002}).

\bibitem[{\citenamefont{Zi\'{n} et~al.}(2005)\citenamefont{Zi\'{n},
  Chwede\'{n}czuk, Veitia, Rz\c{a}\.{z}ewski, and Trippenbach}}]{zin:05}
\bibinfo{author}{\bibfnamefont{P.}~\bibnamefont{Zi\'{n}}},
  \bibinfo{author}{\bibfnamefont{J.}~\bibnamefont{Chwede\'{n}czuk}},
  \bibinfo{author}{\bibfnamefont{A.}~\bibnamefont{Veitia}},
  \bibinfo{author}{\bibfnamefont{K.}~\bibnamefont{Rz\c{a}\.{z}ewski}},
  \bibnamefont{and}
  \bibinfo{author}{\bibfnamefont{M.}~\bibnamefont{Trippenbach}},
  \bibinfo{journal}{Phys. Rev. Lett.} \textbf{\bibinfo{volume}{94}},
  \bibinfo{pages}{200401} (\bibinfo{year}{2005}).

\bibitem[{\citenamefont{Zi\'{n} et~al.}(2006)\citenamefont{Zi\'{n},
  Chwede\'{n}czuk, and Trippenbach}}]{zin:06}
\bibinfo{author}{\bibfnamefont{P.}~\bibnamefont{Zi\'{n}}},
  \bibinfo{author}{\bibnamefont{Chwede\'{n}czuk}}, \bibnamefont{and}
  \bibinfo{author}{\bibfnamefont{M.}~\bibnamefont{Trippenbach}},
  \bibinfo{journal}{Phys. Rev. A} \textbf{\bibinfo{volume}{73}},
  \bibinfo{eid}{033602} (\bibinfo{year}{2006}).

\bibitem[{\citenamefont{Norrie et~al.}(2005)\citenamefont{Norrie, Ballagh, and
  Gardiner}}]{norrie:05}
\bibinfo{author}{\bibfnamefont{A.~A.} \bibnamefont{Norrie}},
  \bibinfo{author}{\bibfnamefont{R.~J.} \bibnamefont{Ballagh}},
  \bibnamefont{and} \bibinfo{author}{\bibfnamefont{C.~W.}
  \bibnamefont{Gardiner}}, \bibinfo{journal}{Phys. Rev. Lett.}
  \textbf{\bibinfo{volume}{94}}, \bibinfo{eid}{040401}
  (\bibinfo{year}{2005}).

\bibitem[{\citenamefont{Norrie et~al.}(2006)\citenamefont{Norrie, Ballagh, and
  Gardiner}}]{norrie:06}
\bibinfo{author}{\bibfnamefont{A.~A.} \bibnamefont{Norrie}},
  \bibinfo{author}{\bibfnamefont{R.~J.} \bibnamefont{Ballagh}},
  \bibnamefont{and} \bibinfo{author}{\bibfnamefont{C.~W.}
  \bibnamefont{Gardiner}}, \bibinfo{journal}{Phys. Rev. A}
  \textbf{\bibinfo{volume}{73}}, \bibinfo{pages}{043617}
  (\bibinfo{year}{2006}).

\bibitem[{\citenamefont{Savage et~al.}(2006)\citenamefont{Savage, Schwenn, and
  Kheruntsyan}}]{savage:06}
\bibinfo{author}{\bibfnamefont{C.~M.} \bibnamefont{Savage}},
  \bibinfo{author}{\bibfnamefont{P.~E.} \bibnamefont{Schwenn}},
  \bibnamefont{and} \bibinfo{author}{\bibfnamefont{K.~V.}
  \bibnamefont{Kheruntsyan}}, \bibinfo{journal}{Phys. Rev. A}
  \textbf{\bibinfo{volume}{74}}, \bibinfo{pages}{033620}
  (\bibinfo{year}{2006}).

\bibitem[{\citenamefont{Deuar and Drummond}(2007)}]{deuar:07}
\bibinfo{author}{\bibfnamefont{P.}~\bibnamefont{Deuar}} \bibnamefont{and}
  \bibinfo{author}{\bibfnamefont{P.~D.} \bibnamefont{Drummond}},
  \bibinfo{journal}{Phys. Rev. Lett.} \textbf{\bibinfo{volume}{98}},
  \bibinfo{pages}{120402} (\bibinfo{year}{2007}).

\bibitem[{\citenamefont{Perrin et~al.}(2008)\citenamefont{Perrin, Savage,
  Boiron, Krachmalnicoff, Westbrook, and Kheruntsyan}}]{perrin:08}
\bibinfo{author}{\bibfnamefont{A.}~\bibnamefont{Perrin}},
  \bibinfo{author}{\bibfnamefont{C.~M.} \bibnamefont{Savage}},
  \bibinfo{author}{\bibfnamefont{D.}~\bibnamefont{Boiron}},
  \bibinfo{author}{\bibfnamefont{V.}~\bibnamefont{Krachmalnicoff}},
  \bibinfo{author}{\bibfnamefont{C.~I.} \bibnamefont{Westbrook}},
  \bibnamefont{and} \bibinfo{author}{\bibfnamefont{K.~V.}
  \bibnamefont{Kheruntsyan}}, \bibinfo{journal}{New J. Phys.}
  \textbf{\bibinfo{volume}{10}}, \bibinfo{pages}{045021}
  (\bibinfo{year}{2008}).

\bibitem[{\citenamefont{M\o{}lmer et~al.}(2008)\citenamefont{M\o{}lmer, Perrin,
  Krachmalnicoff, Leung, Boiron, Aspect, and Westbrook}}]{molmer:07}
\bibinfo{author}{\bibfnamefont{K.}~\bibnamefont{M\o{}lmer}},
  \bibinfo{author}{\bibfnamefont{A.}~\bibnamefont{Perrin}},
  \bibinfo{author}{\bibfnamefont{V.}~\bibnamefont{Krachmalnicoff}},
  \bibinfo{author}{\bibfnamefont{V.}~\bibnamefont{Leung}},
  \bibinfo{author}{\bibfnamefont{D.}~\bibnamefont{Boiron}},
  \bibinfo{author}{\bibfnamefont{A.}~\bibnamefont{Aspect}}, \bibnamefont{and}
  \bibinfo{author}{\bibfnamefont{C.~I.} \bibnamefont{Westbrook}},
  \bibinfo{journal}{Phys. Rev. A} \textbf{\bibinfo{volume}{77}},
  \bibinfo{pages}{033601} (\bibinfo{year}{2008}),

\bibitem[{\citenamefont{Chikkatur et~al.}(2000)\citenamefont{Chikkatur,
  G\"{o}rlitz, Stamper-Kurn, Inouye, Gupta, and Ketterle}}]{chikkatur:00}
\bibinfo{author}{\bibfnamefont{A.}~\bibnamefont{Chikkatur}},
  \bibinfo{author}{\bibfnamefont{A.}~\bibnamefont{G\"{o}rlitz}},
  \bibinfo{author}{\bibfnamefont{D.}~\bibnamefont{Stamper-Kurn}},
  \bibinfo{author}{\bibfnamefont{S.}~\bibnamefont{Inouye}},
  \bibinfo{author}{\bibfnamefont{S.}~\bibnamefont{Gupta}}, \bibnamefont{and}
  \bibinfo{author}{\bibfnamefont{W.}~\bibnamefont{Ketterle}},
  \bibinfo{journal}{Phys. Rev. Lett.} \textbf{\bibinfo{volume}{85}},
  \bibinfo{pages}{483} (\bibinfo{year}{2000}).

\bibitem[{\citenamefont{Gibble et~al.}(1995)\citenamefont{Gibble, Chang, and
  Legere}}]{gibble:95}
\bibinfo{author}{\bibfnamefont{K.}~\bibnamefont{Gibble}},
  \bibinfo{author}{\bibfnamefont{S.}~\bibnamefont{Chang}}, \bibnamefont{and}
  \bibinfo{author}{\bibfnamefont{R.}~\bibnamefont{Legere}},
  \bibinfo{journal}{Phys. Rev. Lett.} \textbf{\bibinfo{volume}{75}},
  \bibinfo{pages}{2666} (\bibinfo{year}{1995}).

\bibitem[{\citenamefont{Katz et~al.}(2005)\citenamefont{Katz, Rowen, Ozeri, and
  Davidson}}]{katz:05}
\bibinfo{author}{\bibfnamefont{N.}~\bibnamefont{Katz}},
  \bibinfo{author}{\bibfnamefont{E.}~\bibnamefont{Rowen}},
  \bibinfo{author}{\bibfnamefont{R.}~\bibnamefont{Ozeri}}, \bibnamefont{and}
  \bibinfo{author}{\bibfnamefont{N.}~\bibnamefont{Davidson}},
  \bibinfo{journal}{Phys. Rev. Lett.} \textbf{\bibinfo{volume}{95}},
  \bibinfo{pages}{220403} (\bibinfo{year}{2005}).

\bibitem[{\citenamefont{Schellekens et~al.}(2005)\citenamefont{Schellekens,
  Hoppeler, Perrin, Viana~Gomes, Boiron, Westbrook, and
  Aspect}}]{schellekens:05}
\bibinfo{author}{\bibfnamefont{M.}~\bibnamefont{Schellekens}},
  \bibinfo{author}{\bibfnamefont{R.}~\bibnamefont{Hoppeler}},
  \bibinfo{author}{\bibfnamefont{A.}~\bibnamefont{Perrin}},
  \bibinfo{author}{\bibfnamefont{J.}~\bibnamefont{Viana~Gomes}},
  \bibinfo{author}{\bibfnamefont{D.}~\bibnamefont{Boiron}},
  \bibinfo{author}{\bibfnamefont{C.~I.} \bibnamefont{Westbrook}},
  \bibnamefont{and} \bibinfo{author}{\bibfnamefont{A.}~\bibnamefont{Aspect}},
  \bibinfo{journal}{Science} \textbf{\bibinfo{volume}{310}},
  \bibinfo{pages}{648} (\bibinfo{year}{2005}).

\bibitem[{\citenamefont{Jeltes et~al.}(2007)\citenamefont{Jeltes, McNamara,
  Hogervorst, Vassen, Krachmalnicoff, Schellekens, Perrin, Chang, Boiron,
  Aspect et~al.}}]{jeltes:07}
\bibinfo{author}{\bibfnamefont{T.}~\bibnamefont{Jeltes}},
  \bibinfo{author}{\bibfnamefont{J.~M.} \bibnamefont{McNamara}},
  \bibinfo{author}{\bibfnamefont{W.}~\bibnamefont{Hogervorst}},
  \bibinfo{author}{\bibfnamefont{W.}~\bibnamefont{Vassen}},
  \bibinfo{author}{\bibfnamefont{V.}~\bibnamefont{Krachmalnicoff}},
  \bibinfo{author}{\bibfnamefont{M.}~\bibnamefont{Schellekens}},
  \bibinfo{author}{\bibfnamefont{A.}~\bibnamefont{Perrin}},
  \bibinfo{author}{\bibfnamefont{H.}~\bibnamefont{Chang}},
  \bibinfo{author}{\bibfnamefont{D.}~\bibnamefont{Boiron}},
  \bibinfo{author}{\bibfnamefont{A.}~\bibnamefont{Aspect}},
  \bibnamefont{et~al.}, \bibinfo{journal}{Nature}
  \textbf{\bibinfo{volume}{445}}, \bibinfo{pages}{402} (\bibinfo{year}{2007}).

\bibitem[{\citenamefont{Scully and Zubairy}(1997)}]{ScullyQO}
\bibinfo{author}{\bibfnamefont{M.~O.} \bibnamefont{Scully}} \bibnamefont{and}
  \bibinfo{author}{\bibfnamefont{M.~S.} \bibnamefont{Zubairy}},
  \emph{\bibinfo{title}{Quantum Optics}} (\bibinfo{publisher}{Cambridge
  University Press; 1 edition (September 28, 1997)}, \bibinfo{year}{1997}).

\bibitem[{\citenamefont{Lifshitz and Pitaevskii}(1980)}]{Lif:80}
\bibinfo{author}{\bibfnamefont{E.~M.} \bibnamefont{Lifshitz}} \bibnamefont{and}
  \bibinfo{author}{\bibfnamefont{L.~P.} \bibnamefont{Pitaevskii}},
  \emph{\bibinfo{title}{Statistical Physics, Part 2}}
  (\bibinfo{publisher}{Pergamon Press, Oxford}, \bibinfo{year}{1980}).

\bibitem[{\citenamefont{\"{O}hberg et~al.}(1997)\citenamefont{\"{O}hberg,
  Surkov, Tittonen, Stenholm, Wilkens, and Shlyapnikov}}]{Ohberg:97}
\bibinfo{author}{\bibfnamefont{P.}~\bibnamefont{\"{O}hberg}},
  \bibinfo{author}{\bibfnamefont{E.~L.} \bibnamefont{Surkov}},
  \bibinfo{author}{\bibfnamefont{I.}~\bibnamefont{Tittonen}},
  \bibinfo{author}{\bibfnamefont{S.}~\bibnamefont{Stenholm}},
  \bibinfo{author}{\bibfnamefont{M.}~\bibnamefont{Wilkens}}, \bibnamefont{and}
  \bibinfo{author}{\bibfnamefont{G.~V.} \bibnamefont{Shlyapnikov}},
  \bibinfo{journal}{Phys. Rev. A} \textbf{\bibinfo{volume}{56}},
  \bibinfo{pages}{R3346} (\bibinfo{year}{1997}).

\bibitem[{\citenamefont{Lisa et~al.}(2005)\citenamefont{Lisa, Pratt, Stoltz,
  and Wiedemann}}]{lisa:05}
\bibinfo{author}{\bibfnamefont{M.}~\bibnamefont{Lisa}},
  \bibinfo{author}{\bibfnamefont{S.}~\bibnamefont{Pratt}},
  \bibinfo{author}{\bibfnamefont{R.}~\bibnamefont{Stoltz}}, \bibnamefont{and}
  \bibinfo{author}{\bibfnamefont{U.}~\bibnamefont{Wiedemann}},
  \bibinfo{journal}{Ann. Rev. Nucl. Part.Sci.} \textbf{\bibinfo{volume}{55}},
  \bibinfo{pages}{357} (\bibinfo{year}{2005}).

\bibitem[{\citenamefont{Wong and Zhang}(2007)}]{wong:07}
\bibinfo{author}{\bibfnamefont{C.-Y.} \bibnamefont{Wong}} \bibnamefont{and}
  \bibinfo{author}{\bibfnamefont{W.-N.} \bibnamefont{Zhang}},
  \bibinfo{journal}{Phys. Rev. C} \textbf{\bibinfo{volume}{76}},
  \bibinfo{pages}{034905} (\bibinfo{year}{2007}).

\end{thebibliography}

\end{document}